\documentclass{article}

\pdfoutput=1

\usepackage[preprint,nonatbib]{neurips_2024}

\usepackage[utf8]{inputenc} 
\usepackage[T1]{fontenc}    
\usepackage{hyperref}       
\usepackage{url}            
\usepackage{booktabs}       
\usepackage{amsfonts}       
\usepackage{nicefrac}       
\usepackage{microtype}      
\usepackage{xcolor}         
\usepackage{graphicx}
\usepackage{xspace}
\usepackage{multirow}
\usepackage{tcolorbox}
\usepackage{caption}
\usepackage{listings}

\definecolor{lightgray}{rgb}{.9,.9,.9}
\definecolor{darkgray}{rgb}{.4,.4,.4}
\definecolor{purple}{rgb}{0.65, 0.12, 0.82}

\lstdefinelanguage{JavaScript}{
  keywords={typeof, new, true, false, catch, function, return, null, catch, switch, var, if, in, while, do, else, case, break},
  keywordstyle=\color{blue}\bfseries,
  ndkeywords={class, export, boolean, throw, implements, import, this},
  ndkeywordstyle=\color{darkgray}\bfseries,
  identifierstyle=\color{black},
  sensitive=false,
  comment=[l]{//},
  morecomment=[s]{/*}{*/},
  commentstyle=\color{purple}\ttfamily,
  stringstyle=\color{red}\ttfamily,
  morestring=[b]',
  morestring=[b]"
}

\newcommand{\sys}{\texttt{JiT-Codegen}\xspace}

\usepackage{amssymb}
\usepackage{pifont}
\newcommand{\xmark}{\ding{55}}%

\title{Every Software as an Agent: Blueprint and Case Study}

\author{%
  Mengwei Xu \\
  Beiing University of Posts and Telecommunications \\
  \texttt{mwx@bupt.edu.cn} \\
}

\begin{document}

\maketitle

\begin{abstract}
The rise of (multimodal) large language models (LLMs) has shed light on software agent -- where software can understand and follow user instructions in natural language.
However, existing approaches such as API-based and GUI-based agents are far from satisfactory at accuracy and efficiency aspects.
Instead, we advocate to endow LLMs with access to the software internals (source code and runtime context) and the permission to dynamically inject generated code into software for execution.
In such a whitebox setting, one may better leverage the software context and the coding ability of LLMs.
We then present an overall design architecture and case studies on two popular web-based desktop applications.
We also give in-depth discussion of the challenges and future directions.
We deem that such a new paradigm has the potential to fundamentally overturn the existing software agent design, and finally  creating a digital world in which software can comprehend, operate, collaborate, and even think to meet complex user needs.
\end{abstract}

\section{Introduction}

\begin{figure}[h]
    \centering
    \includegraphics[width=0.98\textwidth]{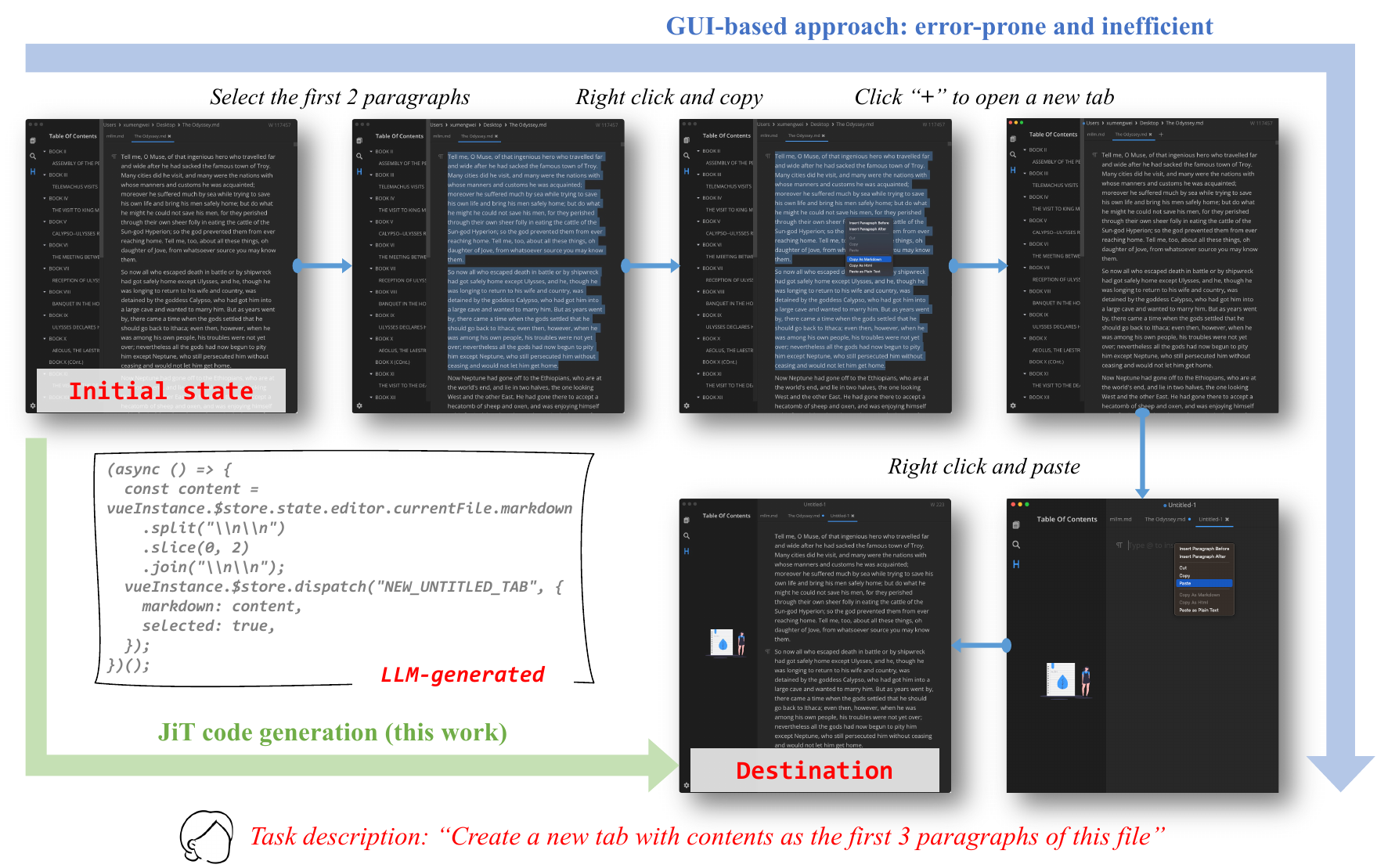}
    \caption{An illustrative example showing the difference between GUI-based agent and our proposal of \sys-based agent. The software tested is a popular open-sourced markdown editor.}
    \label{fig:example}
\end{figure}

The complexity of modern software, such as mobile and desktop applications, often hampers users’ ability to interact with these systems efficiently, particularly for individuals with cognitive challenges. This complexity is manifested in the increasing number of features, the density of graphical user interface (GUI) elements within a window, and the depth of action sequences required to complete tasks. Even when users are capable of operating the software, external factors such as driving can restrict their ability to interact with the system.

To enhance software accessibility at all times and in any context, various approaches have leveraged machine learning-powered software agents~\cite{li2024personal}\footnote{This work targets turning into agent every software, not just user applications with GUI.}. The ultimate objective is to enable software to understand natural language instructions and perform tasks autonomously, akin to human behavior. This goal has become increasingly viable with the advent of large language models (LLMs), which are adept at natural language understanding, reasoning, and planning. Broadly, there are two primary paths for creating software agents: API-based and GUI-based agents.

API-based agents~\cite{xie2024droidcall,erdogan2024tinyagent} necessitate that software developers pre-define API functions that map to user instructions. However, this approach suffers from limited scalability and flexibility, making it ill-suited for handling arbitrary user commands, and thus does not lend itself to the creation of a general software agent. On the other hand, GUI-based agents~\cite{wen2024autodroid,gao2024mobileviews,zhang2024large} -- by mimicking human understanding and actions on graphical interfaces -- have recently been identified as a promising direction towards more generalized software agents. While multimodal LLMs have shown impressive performance in GUI understanding, the success rate of end-to-end task completion remains low, due to error accumulation across multi-step GUI interactions. Additionally, task completion times are often prolonged by the need for multiple rounds of rendering and LLM invocations.

Both approaches, however, assume that LLMs have no access to the software’s internal workings. We argue that this assumption presents a fundamental barrier to building a truly powerful, efficient, and general software agent. In response, this work proposes granting LLMs access to the software’s internals -- specifically, the complete source code, documentation, and the ability to inject generated code for real-time execution within the software runtime.

This paper introduces a novel approach to software-as-agent: just-in-time code generation and in-software execution (denoted as \sys). Drawing an analogy to just-in-time (JIT) compilation, \sys enables an agent to generate \textit{action code} that directly interacts with the software’s runtime context such as functions, data structures, databases, and UI elements. By providing the agent with full access to the software’s source code, the LLM can translate a user’s natural language instruction into executable code that operates within the runtime environment. While previous research has explored software functionality understanding through offline self-exploration and leveraging LLMs for runtime software interaction~\cite{wen2024autodroidv2,lu2024turn,lee2024mobilegpt}, \sys represents the first attempt to have an LLM generate and execute code within the software runtime itself. Figure~\ref{fig:example} illustrates an example where a \sys agent efficiently addresses a complex task involving five GUI interactions with just two lines of code.

\begin{figure}
    \centering
     \includegraphics[width=0.9\textwidth]{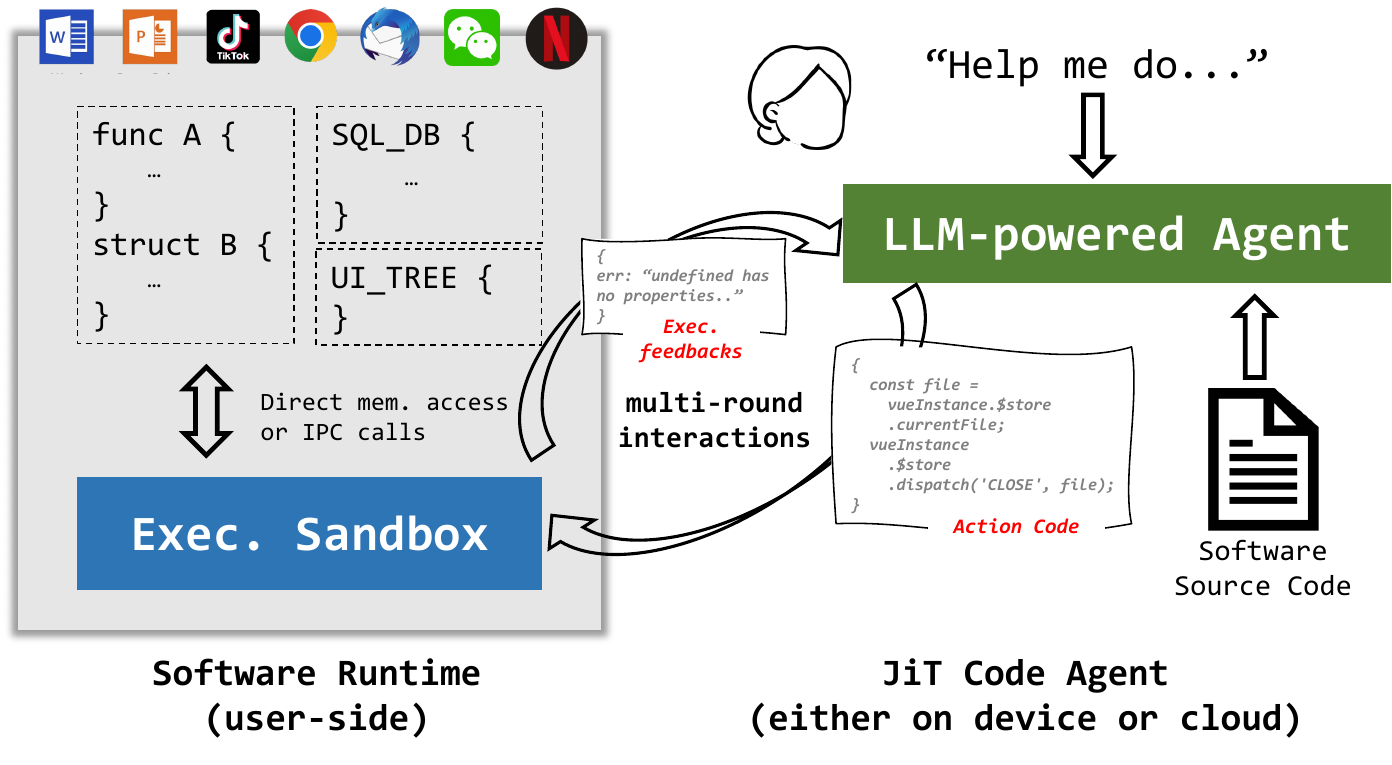}
    \caption{An overview of software agent using \sys.}
    \label{fig:overview}
\end{figure}


Notably, \sys is \textit{not} mutually exclusive to API-based or GUI-based agents; rather, these approaches can complement one another. For example, \sys can be seen as an on-demand API-based solution, where a function is dynamically created at runtime based on the task at hand. This approach prioritizes flexibility over certainty. Furthermore, once a generated action code successfully completes a task, it can be encapsulated into an API for future invocation. Additionally, \sys can leverage GUI understanding and manipulation at the code level, recognizing that GUIs are ultimately rendered and controlled through source code (e.g., Android's \texttt{View} and \texttt{OnClickListener}).

This paper serves as a blueprint for \sys agents, presenting an overall design architecture, case studies utilizing web-based PC applications, and an in-depth discussion of the challenges and future directions. Our long-term vision is to transform all software into agents, creating a multi-agent system in which software works collaboratively to meet user needs in a more intelligent and dynamic manner --  a digital world where software comprehends, actions, and cooperates as humans.


\section{The Status Quo and Their Inadequacies}

While the efforts to enable natural language communication between users and software date back to the 1960s~\cite{shoebox}, the recent emergence of large language models (LLMs) has significantly advanced the practical application of this concept. This section provides an overview of the current practice to build software agents.

\textbf{API-based Approach}.
The most direct method to facilitate natural language interactions with software is through pre-defined APIs (often referred to as intents), created by developers for key software functions. Prior to the advent of LLMs, classification models were employed to map user instructions to one of several pre-defined API calls, sometimes with associated parameters. The zero-shot language understanding capabilities of LLMs now enable more precise identification of APIs and parameters, as well as dynamic function registration~\cite{xie2024droidcall, erdogan2024tinyagent}. However, the API-based approach has a notable limitation: it depends on developers to define functions, which restricts its ability to address more general or open-ended user tasks in a dynamic fashion.

\textbf{GUI-based Approach}.
The GUI-based approach mimics human-device interaction by interpreting and manipulating device screen elements or pixels~\cite{zhang2024large}, as demonstrated in applications like Claude Computer Use~\cite{computer-use} and OpenAI Operator~\cite{openai-operator}. This approach has garnered significant attention in both academic and industrial circles due to the success of visual language models (VLMs). Unlike the API-based method, the GUI-based approach can, in theory, address any task that users can perform, as both approaches operate at the same granularity (UI elements). Nevertheless, practical challenges remain, such as the inaccuracy of VLMs in grounding complex visual objects. Furthermore, GUI-based operations often involve multiple steps, each requiring screen re-rendering and understanding. Errors accumulating at each step can ultimately lead to task failure.

\textbf{GUI-to-API Approach}.
A hybrid model explores software GUI states offline to extract APIs for online use. Upon receiving a user task, the agent first employs an LLM to map the instruction to the corresponding function, then executes it through the underlying GUI operations. This approach treats GUI operations as function calls. For example, AXIS~\cite{lu2024turn} uses LLM-based agents to explore the software environment via a unified interface to gather runtime knowledge, which is then consolidated into APIs for online invocation.
AutoDroid-v2~\cite{wen2024autodroidv2} leverages the coding capabilities of small language models to generate UI operation scripts for online execution. Although this hybrid approach balances ubiquity and efficiency, it is constrained by two primary drawbacks: (1) the need for a cumbersome offline GUI exploration phase, and (2) difficulty in managing tasks with high GUI dynamics.

All existing methods are ``non-intrusive'', assuming that the LLM has no access to the software's runtime environment and cannot influence its execution. We deem such an assumption fundamentally constrain the agent's ability to perform complex user tasks. On both desktop and mobile application benchmarks~\cite{zhang2024llamatouch, xie2024osworld, rawles2024androidworld}, current agents exhibit low task completion rates (typically under 20\%).
Even worse, many of real-world complex user tasks can be hardly addressed through GUI, e.g., ``tell me where I am going to date today''. In this case, the GUI-agent may resort to the searching functionality in an email/IMA/etc apps, yet the searching query is hard to define correctly and clearly.

Unlike any of above approaches, this work proposes a new paradigm where the agent gains access to the software's internals, as outlined in the following section.

\section{A Vision of Software Agent}

We propose a fundamentally distinct approach for software agents: \texttt{just-in-time code generation and in-software execution}. As illustrated in Figure~\ref{fig:overview}, a \sys-based software agent comprises two core components: an \textbf{LLM-powered Code Agent}, which generates ``action code'' based on user instructions, and an \textbf{Execution Sandbox}, responsible for executing the generated action code within the software's runtime context. A key distinguishing feature of this design, compared to existing code-generation-based agents~\cite{wen2024autodroidv2, lee2024mobilegpt}, is that it injects the generated code directly into the software runtime, enabling it to interact with the runtime's rich context to accomplish complex and pervasive tasks. This runtime context includes in-memory data structures, code, open files, databases, and UI elements.
The design is inspired by the Linux kernel's eBPF subsystem, which enables user programs to inject code that executes in kernel mode and access certain kernel data structures to a limited extent.

The interaction between \texttt{CodeAgent} and \texttt{Sandbox} is bi-directional and iterative.
By bi-directional, \texttt{CodeAgent} sends action code to \texttt{Sandbox} for on-demand execution, and in turn, the \texttt{Sandbox} provides valuable feedbacks to the \texttt{CodeAgent} to refine the generated code through multi-step, more precise iterations.
This feedback could include error messages from the language runtime (e.g., JVM or JavaScript engine) or execution results that inform subsequent rounds of code generation.
The necessity of such multi-round interactions will be demonstrated in Section~\ref{sec:challenges}.
In essence, the \texttt{Sandbox} functions as an external tool through which the \texttt{CodeAgent} interacts with the runtime.

\sys-based agents assume full access to the software's source code and should be deployed by developers. This is reasonable, as developers have strong incentives to make their software more accessible and "agentic" for end-users.

The design of \texttt{CodeAgent} is critical to the overall performance of agent.
To render the approach practical, it must generate \textit{accurate, safe} action code in an \textit{efficient} manner.
Accuracy refers to the agent's ability to successfully complete user tasks, while safety ensures that even if the action code fails to accomplish the intended task, it does not cause harmful or irreversible consequences, such as corrupting a local database. Section~\ref{sec:challenges} will delve deeper into these two aspects. Efficiency concerns the time required for the Code Agent to generate action code, ensuring that the user experiences minimal latency. There has been extensive research aimed at improving the efficiency of LLMs and agent systems~\cite{xu2024survey}.

The design of \texttt{Sandbox} must strike a balance between \textit{programming flexibility, safety}, and \textit{developer efforts}.
Through the \texttt{Sandbox}, the generated action code needs to access necessary in-memory data structures and functions to accomplish user tasks, either through direct memory access (if both in the same process) or inter-process communications.
Simultaneously, the \texttt{Sandbox} must enforce safety constraints on the action code, ensuring that only safe operations are performed by adhering to developer-defined safety rules. To optimize both flexibility and safety, additional effort from developers may be required, such as defining safety rules or exposing IPC mechanisms across processes. The specific implementation of the \texttt{Sandbox} will also depend on the programming language or framework used to develop the software.

\begin{figure}
    \centering
     \includegraphics[width=0.95\textwidth]{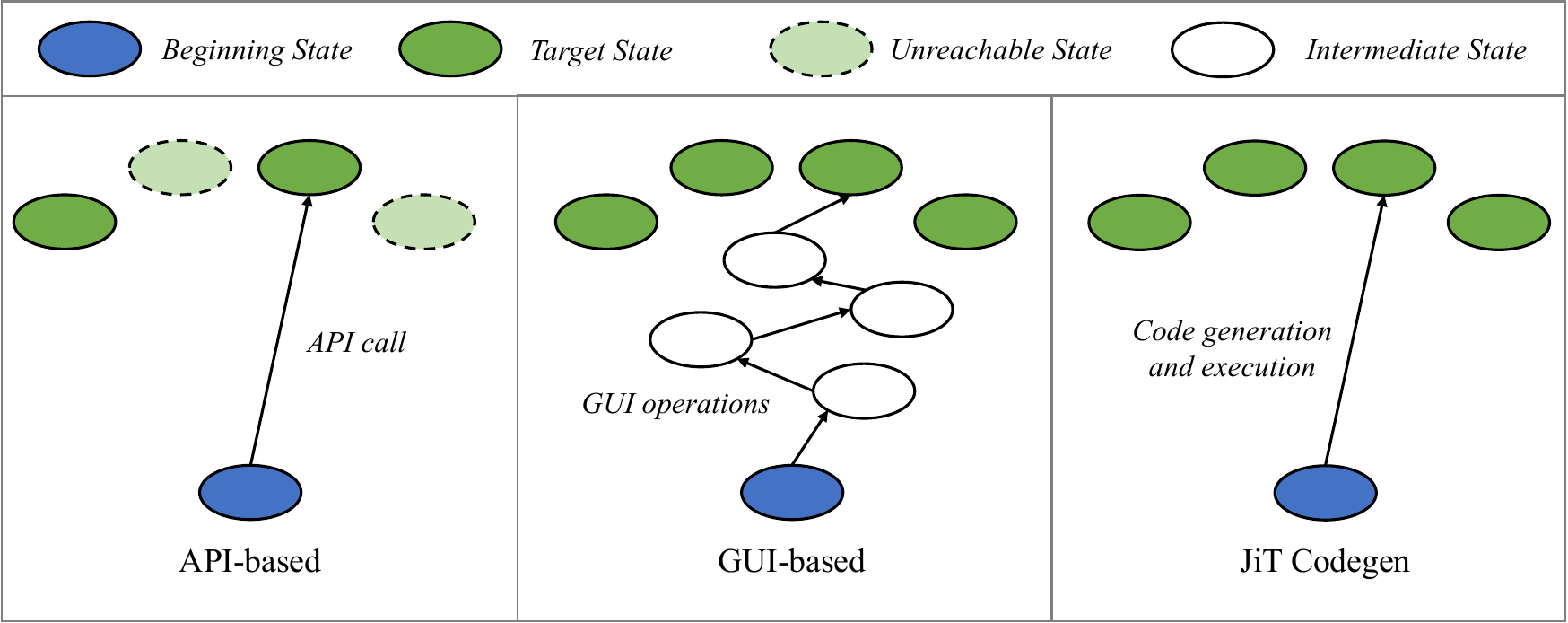}
    \caption{A comparison of software agents from the perspective of state machines.}
    \label{fig:states}
\end{figure}

\begin{table}[]
\caption{A high-level comparison of three approaches to making existing software into an agent. API-based approaches.
Accuracy indicates if an agent can successfully solves a human task; Ubiquity indicates how generalized the tasks can be; Efficiency indicates the runtime cost (e.g., time to completion); Safety indicates if the agent could do harmful things to the system; Invisible Tasks indicate if an agent can solve tasks that do not associate with GUI; Blackbox S/W indicates if the agent can work with software without its source code.}
\label{tab:compare}
\resizebox{\columnwidth}{!}{%
\begin{tabular}{lcccccc}
\hline
& Accuracy & Ubiquity & Efficiency & Safety & Invisible Tasks & Blackbox S/W \\ \hline
\textbf{API-based} & \checkmark & \xmark & \checkmark & \checkmark & \checkmark & \xmark \\ \hline
\textbf{GUI-based} & \xmark & \checkmark & \xmark & \xmark & \xmark & \checkmark \\ \hline
\begin{tabular}[c]{@{}l@{}}\textbf{Source Code Level}\\ \textbf{Generation (Ours)}\end{tabular} & \checkmark & \checkmark\checkmark & \checkmark & \xmark\xmark & \checkmark & \xmark \\ \hline
\end{tabular}%
}
\end{table}

Figure~\ref{fig:states} provides a high-level comparison of different types of software agents, treating the software runtime as a state machine.
API-based agents directly transits from the beginning state to the completion state, but with some states unreachable (not pre-defined);
GUI-based agents can reach a broader set of states through multi-step interactions, where each step leads to an intermediate state.
\sys-based agents allow direct transitions to any target state, effectively enabling a one-step transition\footnote{While one could argue that one API call or action code execution involves many machine-level instructions and transitions between machine states, their complexity is far less than the transitions between GUI states. Thus, we treat them as a single transition.}. Table~\ref{tab:compare} further contrasts these agents along multiple dimensions.
The advantages of \sys-based agents lie in their accuracy (LLMs excel at code understanding and generation), ubiquity (nearly any task can be achieved through code), efficiency (especially in contrast to the step-by-step GUI interpretation and rendering process), and ability to handle tasks that lack a GUI representation. The primary challenge remains ensuring safety, given the potential for hallucinations in LLMs and the significant flexibility they possess in manipulating the software runtime.

\section{Case Studies of Desktop Applications}

\begin{table}[]
\centering
\small
\caption{The task completion rate using \sys agents. \checkmark (*) indicates the task is successfully accomplished through more than one step interaction.}
\begin{tabular}{ll|ccccc|}
\hline
\multicolumn{1}{|c|}{\textbf{Software}} & \multicolumn{1}{c|}{\textbf{Task}} & \textbf{\begin{tabular}[c]{@{}c@{}}Claude 3.5\\ Sonnet\end{tabular}} & \textbf{\begin{tabular}[c]{@{}c@{}}Gemini 2.0\\ Flash\end{tabular}} & \textbf{\begin{tabular}[c]{@{}c@{}}GPT\\ 4o\end{tabular}} & \textbf{o1} & \textbf{o3-mini} \\ \hline
\multicolumn{1}{|l|}{\multirow{5}{*}{\begin{tabular}[c]{@{}l@{}}Music\\ Player\end{tabular}}} & Show my favorite songs & \checkmark & \checkmark & \checkmark & \checkmark & \checkmark \\ \cline{2-7} 
\multicolumn{1}{|l|}{} & Show my listening history & X & X & X & X & X \\ \cline{2-7} 
\multicolumn{1}{|l|}{} & Search for "Hotel California" & \checkmark & X & \checkmark & \checkmark & \checkmark \\ \cline{2-7} 
\multicolumn{1}{|l|}{} & Increase the volume slightly & \checkmark & \checkmark & \checkmark (*) & \checkmark & \checkmark (*) \\ \cline{2-7} 
\multicolumn{1}{|l|}{} & Play the next song & \checkmark & \checkmark & \checkmark & \checkmark & \checkmark \\ \hline
\multicolumn{1}{|l|}{\multirow{5}{*}{\begin{tabular}[c]{@{}l@{}}Markdown\\ Editor\end{tabular}}} & Make the second paragraph bold & X & X & X & X & X \\ \cline{2-7} 
\multicolumn{1}{|l|}{} & Increase the font size by 2 & X & \checkmark & \checkmark & \checkmark & \checkmark \\ \cline{2-7} 
\multicolumn{1}{|l|}{} & Open a new tab & \checkmark & \checkmark & \checkmark & \checkmark & \checkmark \\ \cline{2-7} 
\multicolumn{1}{|l|}{} & Close all other tabs & \checkmark & \checkmark & \checkmark & \checkmark & \checkmark \\ \cline{2-7} 
\multicolumn{1}{|l|}{} & \begin{tabular}[c]{@{}l@{}}Create a new tab with contents as\\ the first 3 paragraphs of this file\end{tabular} & X & \checkmark & \checkmark & X & X \\ \hline
\multicolumn{2}{|c|}{Average} & 6/10 & 7/10 & 8/10 & 7/10 & 7/10 \\ \hline
\end{tabular}
\label{tab:acc}
\end{table}

We have implemented a minimal prototype of \sys framework for software built on Electron~\footnote{\url{https://www.electronjs.org/}}, a web-based, cross-platform development framework for desktop apps.
Popular apps built with Electron includes Dropbox, VS Code, Github Desktop, etc.
To demonstrate the capability of \sys, we use two popular open-sourced apps using Electron: marktext~\footnote{\url{https://github.com/marktext/marktext}} (a markdown editor) and YesPlayMusic~\footnote{\url{https://github.com/qier222/YesPlayMusic}} (a music player).
Both projects receive more than 30K stars in Github and have been widely adored by end users.
We used Copilot for the code searching and Claude 3.5/GPT-4o/o1/o1-mini for code generation\footnote{Since Copilot does not support API-based invocation, we wrote a VSCode plugin to interact with it.}.
To be noted, this is a simplistic, vanilla implementation of \sys agents, with less than 200 LoC in JavaScript and 300 LoC in Python.
We expect it to be a baseline version for future research, e.g., more powerful models and agentic workflow.

As an initial attempt, we selected 5 practical tasks for each app, as shown in Table~\ref{tab:acc}.
We tested the implementation on Apple M2 Pro with MacOS version 13.4.1.

\begin{figure}
    \centering
     \includegraphics[width=0.95\textwidth]{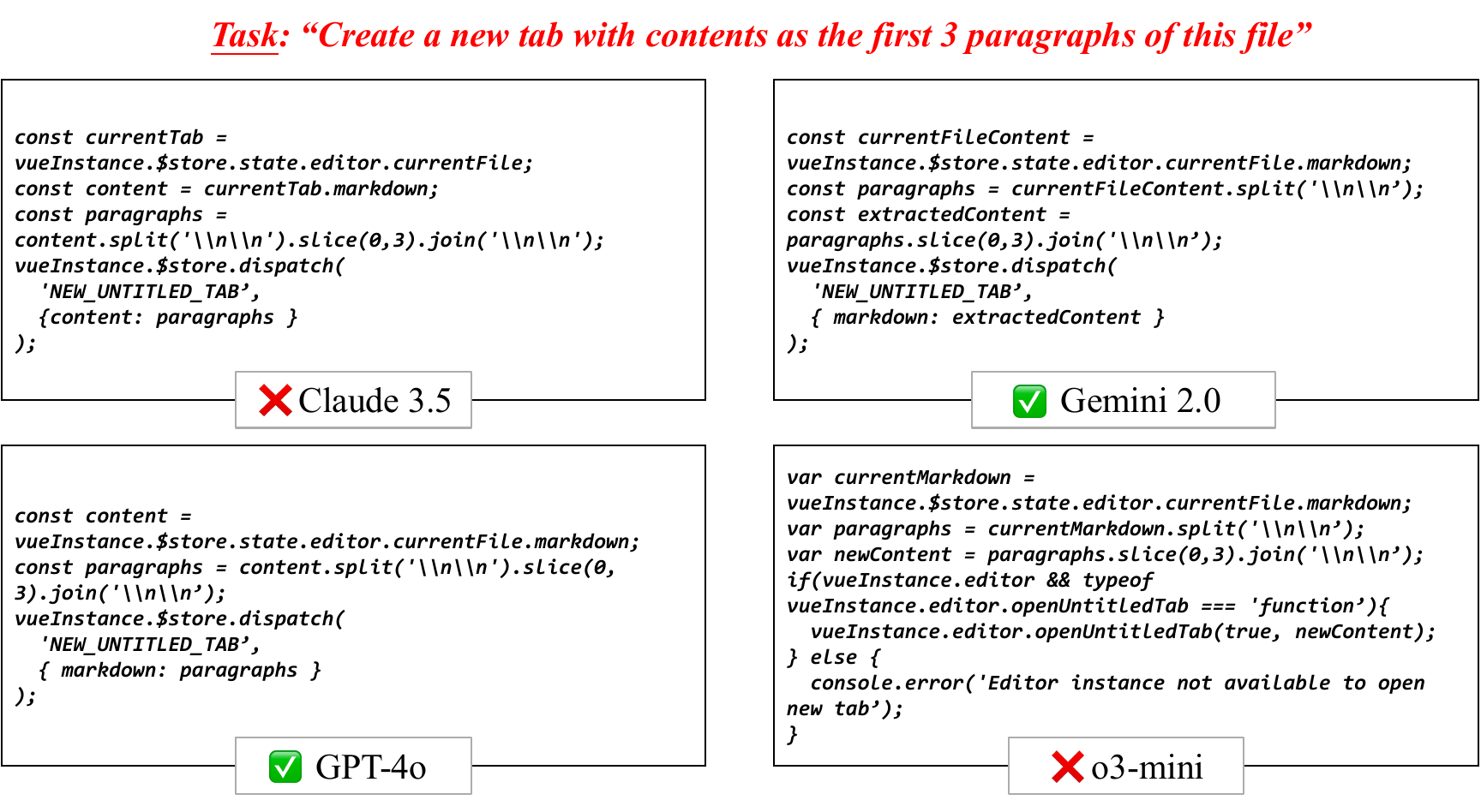}
    \caption{Examples of the action code generated by different LLMs.}
    \label{fig:code-example}
\end{figure}

\begin{lstlisting}[label=code:multi-iterations,caption=An example of how multi-round interactions between execution sandbox and code agent helps task accomplishment. Task: ``increase the volume slightly'' on music player; model: o3-mini.]
// 1st Iteraction (failed): "TypeError: Cannot read property 'volume' of undefined."
let currentVolume = vueInstance.player.volume;
let newVolume = Math.min(currentVolume + 0.1, 1);
vueInstance.player.volume = newVolume;
console.log('Volume increased to', newVolume);

// 2nd Iteraction (failed): "Error: Player component not found."
let player = (vueInstance.$refs && vueInstance.$refs.player) ? vueInstance.$refs.player : vueInstance.player;
if (!player) {
    throw new Error('Player component not found');
}
let currentVolume = player.volume;
let newVolume = Math.min(currentVolume + 0.1, 1);
player.volume = newVolume;
console.log('Volume increased to', newVolume);

// 3rd Interaction. Task successfully accomplished.
let player = vueInstance.player || (window.yesplaymusic && window.yesplaymusic.player) || (vueInstance.$store && vueInstance.$store.state && vueInstance.$store.state.player);
if (!player) {
    throw new Error('Player component not found');
}
let currentVolume = player.volume;
let newVolume = Math.min(currentVolume + 0.1, 1);
player.volume = newVolume;
console.log('Volume increased to', newVolume);
\end{lstlisting}

\textbf{Overall task completion success rate.}
Among the 10 tasks tested, GPT 4o successfully solves 8 tasks with 80\% task completion ratio.
Surprisingly, the reasoning models (o1 and o3-mini) solve only 7 tasks, which is likely attributed to the ineffective prompt engineering.
Figure~\ref{fig:code-example} compares the output from different LLMs on task ``Create a new tab with contents as the first 3 paragraphs of this file''.

\textbf{Feedbacks from execution sandbox helps (sometimes).}
We observe that sometimes the response from the execution sandbox (error messages) can help the code agent to generate the right code through reflection.
An example of 3-round interactions is shown in Listing~\ref{code:multi-iterations}.

\begin{lstlisting}[label=code:DOM,caption=A workable code for task "show my listening history" on the music player app. It requires manipulating the HTML DOM tree. The code is generated by o1 through multi-round prompting.]
vueInstance.$router.push("/library").then(() => {
  // Select all tabs within the tabs container
  const tabs = document.querySelectorAll('.tabs .tab');
  // Check if the Play History tab exists
  // (assuming it's the 6th tab, index 5)
  if (tabs.length >= 6) {
    tabs[5].click();
  } else {
    console.error("Play History tab not found.");
  }
});
\end{lstlisting}

\textbf{An analysis of failed cases.}
For most failed cases, we observe ``salient failure'', where the action code does not successfully accomplish the user task, neither the execution reports any runtime error.
In such cases, the code agent will not continue generating action code, unless the users give extra feedbacks.
For examples, all LLMs encounter salient failure when automating ``show my listening history'' on the music player.
However, we also observe that if we manually keep prompting, the LLM can generate a workable action code eventually, as shown in Listing~\ref{code:DOM}.

\section{Challenges and Future Directions}\label{sec:challenges}

\subsection{Safeguarding against Vulnerable Action Code}

Flexibility comes with cost on safety.
As the action code generated is executed in the software runtime context with access to any data in the memory and disk literally, it could do harmful and irreversible damage to the software as well.
We discuss a few potential safeguarding techniques here.
Notably, the execution sandbox added to software runtime that executes code dynamically downloaded also exposes great security issue which has been characterized by prior work~\cite{yue2009characterizing}.

\textbf{Action code sandboxing.}
One of the most effective ways to prevent unsafe operations is to execute the action code within a restricted environment (sandbox). A sandbox isolates the code from critical system resources (e.g., memory, disk, network), limiting its access to only the necessary parts of the software's runtime. By constraining the execution environment, even if the code is malicious or erroneous, the potential for damage is minimized.

\textbf{Static code analysis and validation.}
Before executing the action code, the agent could run static analysis tools to inspect the code for potential risks such as unsafe memory access, database modification commands, or network access. The goal is to verify that the code adheres to safe practices and does not introduce vulnerabilities. For instance, one could check for unauthorized database writes or calls to sensitive system APIs.
Take a step further, we can leverage LLM-assisted code verifier, i.e., a ``safeguarding agent'', to check the generated action code independently.

\textbf{Audit logs and rollback.}
Another way is to implement real-time monitoring and logging of all actions performed by the action code. By tracking each execution, any unexpected or dangerous behavior can be identified and flagged for review. Audit logs could also allow for a rollback or undo functionality if something goes wrong during execution, particularly for tasks involving sensitive data like database updates.
For tasks that might involve significant risk (e.g., database updates), implement a snapshot or rollback mechanism that saves the state before the code is executed. This allows the system to revert to a known good state if the action produces an unintended or harmful result, reducing the impact of any errors.

\textbf{Rule-enforced LLM generation.}
If we can give concrete ``safety rules'' (e.g., no updates to database), we can enforce LLM to generate code that follows such rules.
If possible, this approach avoids false action code generation, while static/dynamic approaches only detects error but cannot correct it.
Several approaches can potentially achieve such a goal:
prompt engineering with safety rule,
reinforcement learning with safety rules (RLHF),
rule-based post-processing such as filtering,
model fine-tuning with safety-focused synthesis data, etc.
A more advanced direction would be token-level constrained decoding~\cite{beurer2023prompting,beurer2024guiding}.
Nevertheless, formulating ``safety rules'' for \sys would be challenging and often ad-hoc for developers.

All of the approaches discussed involve trade-offs between flexibility, safety, ease of development, and runtime efficiency. The most effective solution is likely to combine several of these methods, striking a balance among these factors based on the specific requirements of developers and users.

\subsection{Improving Action Code Accuracy}

In addition to adopting a more powerful LLM, there have been several potential directions to improve the action code generation accuracy.

\textbf{Codebase-specific agent customization.}
One opportunity is that the codebase of a released software remains mostly static across time, therefore the agent workflow could be fully tuned to it.
For example, one can enhance contextual code understanding through codebase preprocessing.
Rather than retrieving code snippets blindly, preprocess the source code to index key components like functions, classes, variables, and comments. Indexing could be based on semantic understanding, such as function signatures, docstrings, or frequently used design patterns. This enables more precise retrieval of contextually relevant code for the task.
One can also use graph-based representation by treating the software's source code as a graph, where functions and data structures are nodes, and relationships are edges (e.g., function calls or data flow). Use graph structure to retrieve and present the most relevant sections of code based on the user's request, which improves retrieval precision.

Beyond retrieval, one can use the codebase to customize the LLM weights as well.
Through domain-specific fine-tuning the LLM on a dataset of code that reflects the specific codebase, the model better learns nuances, such as preferred coding styles, libraries, and frameworks, resulting in more accurate and contextually appropriate code.
The agent can also utilize a few-shot learning approach where the model is given a set of examples that demonstrate how to translate user instructions into specific actions. This could include common tasks for the software (e.g., updating a database, interacting with an API, etc.), allowing the model to generalize better to new instructions.

\textbf{Agent-friendly programming language and framework.}
A more fundamental approach to improving the performance of software agents involves redesigning the programming language or framework to enhance its comprehensibility and interpretability by LLMs. Traditional programming languages are often designed for human readability and computational efficiency, but they may not align with the way LLMs process and understand code. By tailoring languages to better suit LLMs' capabilities -- such as adopting more consistent syntax, semantic structures, and modularity -- one could significantly improve code generation accuracy and reduce ambiguity. For example, incorporating natural language-like constructs, well-defined and reusable building blocks, and rich metadata annotations could facilitate more seamless integration between the LLM and the software environment. Moreover, optimizing languages for context-aware code generation, such as ensuring explicit runtime information and variable dependencies are easily accessible, would empower LLMs to produce more contextually relevant and functionally accurate code. This rethinking of programming languages not only promises to enhance LLM performance but could also drive the development of more intuitive, human-computer collaborative programming environments, where natural language commands are more directly translatable into executable code.
Notably, such ``programming for agents'' is fundamentally different from ``programming agents''~\cite{khattab2024dspy}.

\textbf{Accurate and timely execution feedbacks}.
As our case study suggests, LLMs can benefit from the feedbacks on the action code execution results (e.g., error messages, task completion or not).
However, most unmanaged programming language like C/Rust do not allow catching runtime error, but instead goes to crash directly.
In such circumstance, how shall the execution sandbox be designed to balance the usability and safety?
Another aspect is on the task completion verification.
The easiest way is human-in-the-loop -- asking users to indicate if the task is successfully completed.
Would it be possible to ask the code agent to generate another piece of ``verification code'' that can check if the action code successfully accomplish the task? This is based on the assumption that generating correct verification code is easier as compared to generating correct action code. We may use two different models (or two fine-tuned variants) to generate two different kinds of code independently.



\subsection{On-Cloud vs On-Devices Codegen}

There has been growing interest in deploying agents and large language models (LLMs) on local devices, such as smartphones and PCs \cite{yi2024phonelm,yuan2024mobile,yin2024llm}. For software agents, on-device LLMs can enhance user privacy by eliminating the need to transmit user instructions to a centralized server, thus preventing the potential leakage of sensitive information. Additionally, on-device processing can reduce end-to-end generation latency, improving the overall responsiveness of the system. However, two significant challenges may hinder the widespread adoption of on-device LLMs for code generation. First, the limited code understanding and generation capabilities of small language models~\cite{lu2024small} restrict their effectiveness in handling complex tasks. Second, on-device code generation requires access to the software's source code, which poses a privacy risk, as this code could potentially be exposed to third parties.

Unlike large-scale models hosted on powerful servers, on-device models are constrained by factors such as memory, processing power, and energy consumption. These limitations prevent them from comprehensively understanding complex code structures and generating highly accurate action code.
Moreover, on-device models may struggle with the context-dependent nature of code generation, where understanding the broader software architecture, interdependencies, and runtime environment is crucial. Potential solutions to this challenge include developing more compact yet efficient LLM architectures that are specifically fine-tuned for on-device applications or making large models affordable on devices through algorithm-system optimizations~\cite{yi2023edgemoe}.
Additionally, the use of specialized models for specific domains (e.g., mobile UI development or database interactions) could improve performance by focusing the model's capabilities on a narrower range of tasks, making it more resource-efficient.

The second challenge is the need for on-device code generation to access the software’s source code, which raises significant privacy and security concerns. For effective code generation, an LLM needs to understand the structure, logic, and dependencies within the software's source code. However, storing and processing the source code locally on the device exposes it to potential risks, particularly if the device is compromised or if third-party applications gain unauthorized access. This situation is especially critical in scenarios where the software handles sensitive user data, such as in banking or healthcare applications, where privacy and security are paramount. One potential solution is to implement more robust access control mechanisms that limit which parts of the source code are accessible to the on-device agent. For example, source code could be encrypted or obfuscated in a way that allows the LLM to generate code without directly exposing sensitive details.

\subsection{Multi-Software Agent Systems}

\begin{figure}[h]
    \centering
    \includegraphics[width=0.9\textwidth]{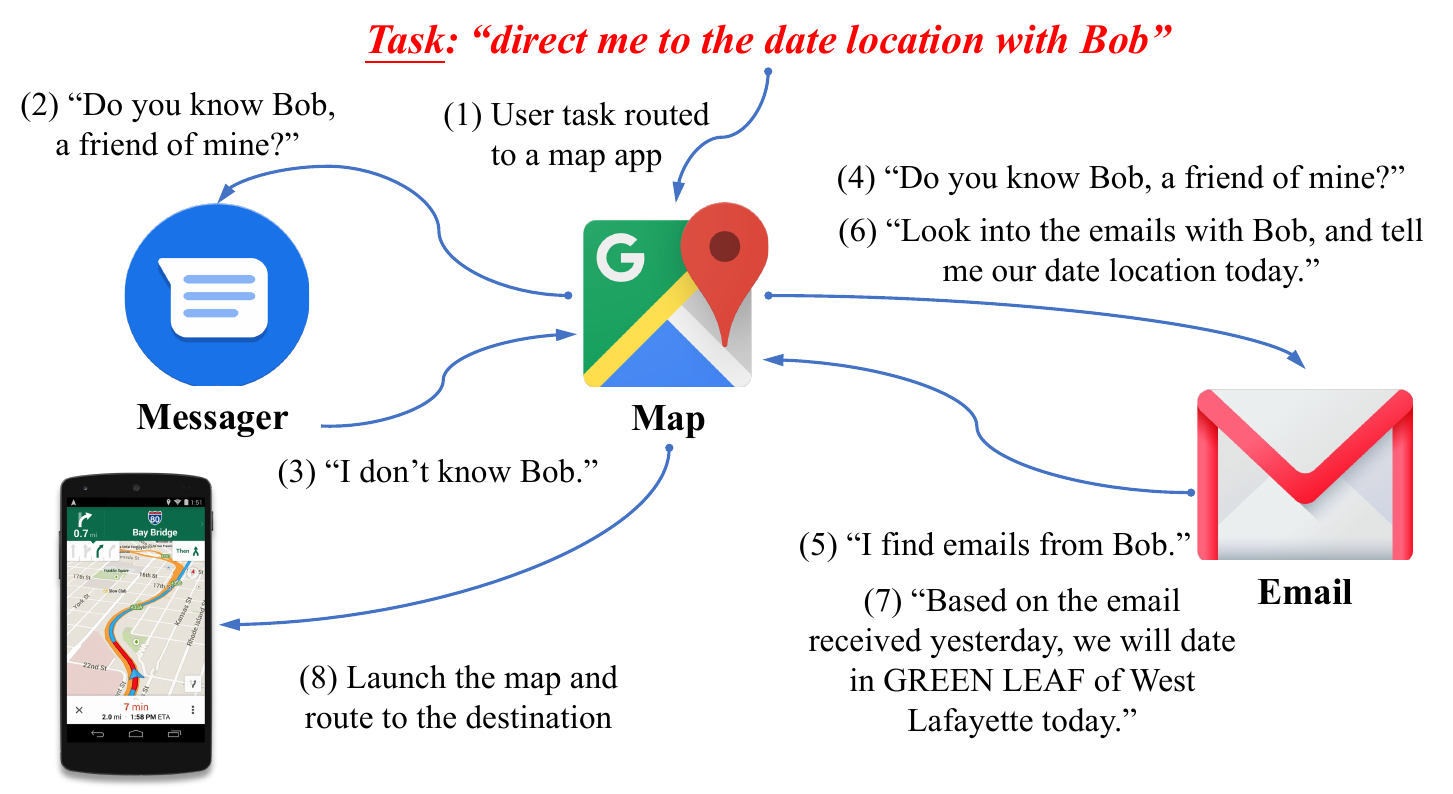}
    \caption{An example showing how multiple software agents collectively solve a user task.}
    \label{fig:multi-agent}
\end{figure}

Once every software can talk and action as humans, a digital world can be established where every software agent behaves as an independent entity and communicates with others in natural language.
Figure~\ref{fig:multi-agent} shows an illustrative of how multiple software agents collaboratively solves a user task.
Such a system differs from existing multi-agent systems~\cite{guo2024large} that every software has rich, independent, and dynamic context (e.g., software source code, documentation, and runtime).
Accordingly, interesting research questions arise, including but not limited to the followings.

\textbf{Scaling to a large number of software agents.}
It is critical to scale up the capability and efficiency of multi-agent system with increased number of software agents involved.
Many aspects of research problems need to be addressed.
For example, the system must take a proper communication topology, among choices of centralized, decentralized, hierarchical, shared message pool, etc.
As the number of agents increases, maintaining performance and responsiveness becomes a challenge. Strategies like load balancing, dynamic agent allocation, and distributed task management can help address these scalability concerns while maintaining efficient resource utilization.
Second, software agents may proactively share/transfer knowledge, i.e., when one agent learns or performs a task, the outcome or insight could be transferred to other agents in the system. For example, an agent that performs a software update could pass knowledge about new APIs or changes in the software to other agents, ensuring uniform knowledge across the system. This approach enables scalability and reduces the need for redundant computations.
Third, for complex user tasks, how to route the task to the proper software agents (groups) is non-trivial as often multiple software agents can potentially solve the same task.
Morever, in a multi-agent environment, failure of one agent should not result in the failure of the entire system. Mechanisms such as agent redundancy, failover protocols, and self-healing systems (where agents can repair or replace faulty agents) could be employed to ensure continuous, reliable operation.

\textbf{Permission Control}
Allowing direct communication and collaboration between software agents introduces significant challenges in permission control, as each agent may have different levels of access to data, functions, and system resources. For instance, a software agent without direct permission to access sensitive data, such as GPS location, could potentially obtain that data through another agent, thereby bypassing established security protocols. This situation resembles permission re-delegation attacks in web browsers and mobile devices, where malicious apps gain unauthorized access by exploiting the permissions granted to other apps~\cite{felt2011permission}.
Therefore, implementing effective permission management is essential to ensure that agents operate within the boundaries set by system administrators or users, preventing unauthorized actions and mitigating security risks. Agents must be restricted to only the permissions necessary to perform their tasks, and these restrictions should be dynamically adaptable based on the context of each interaction. Without proper controls, the interactions between agents could lead to privilege escalation, data leakage, or malicious activity.

However, managing permissions in a multi-agent system is inherently complex. Since agents communicate in natural language, the intentions and scope of permissions are often ambiguous and difficult to interpret. Natural language, by its very nature, lacks the precision required for clear-cut permission definitions, leading to the possibility of miscommunication or unintended access.

\section{Conclusions}

This paper introduces a novel approach to transforming software into intelligent agents through just-in-time code generation and in-software execution. Unlike traditional API-based or GUI-based agents, which are limited by pre-defined functions or static interaction models, \sys empowers software agents with full access to a program’s source code and runtime context. By leveraging large language models (LLMs), \sys enables dynamic, on-the-fly generation of action code that interacts directly with a software’s runtime environment, facilitating seamless and efficient task execution. This work presents a overall design for \sys, illustrating its application in web-based PC applications, and demonstrates how it complements existing agent frameworks, enabling a more flexible, scalable, and powerful approach to software automation.

\bibliographystyle{plain}
\bibliography{ref.bib}

\end{document}